\documentclass{article}
\usepackage{epsfig}

\date{\today}

\usepackage{amsmath}
\usepackage{amssymb}
\usepackage[hang,nooneline,scriptsize]{subfigure}
\begin{document}

\title{Boson stars, neutron stars and black holes  in five dimensions}

\author{{\large Yves Brihaye  \footnote{email: yves.brihaye@umons.ac.be}}$^{\dagger}$ and 
  {\large T\'erence Delsate \footnote{email: terence.delsate@umons.ac.be}}$^{\dagger}$ 
\\ \\
$^{\dagger}${\small Physique-Math\'ematique, Universit\'e de
Mons, 7000 Mons, Belgium}\\ 
}

\date{\today}
\setlength{\footnotesep}{0.5\footnotesep}
\newcommand{\dd}{\mbox{d}}
\newcommand{\tr}{\mbox{tr}}
\newcommand{\la}{\lambda}
\newcommand{\ka}{\kappa}
\newcommand{\f}{\phi}
\newcommand{\vf}{\varphi}
\newcommand{\F}{\Phi}
\newcommand{\al}{\alpha}
\newcommand{\ga}{\gamma}
\newcommand{\de}{\delta}
\newcommand{\si}{\sigma}
\newcommand{\bomega}{\mbox{\boldmath $\omega$}}
\newcommand{\bsi}{\mbox{\boldmath $\sigma$}}
\newcommand{\bchi}{\mbox{\boldmath $\chi$}}
\newcommand{\bal}{\mbox{\boldmath $\alpha$}}
\newcommand{\bpsi}{\mbox{\boldmath $\psi$}}
\newcommand{\brho}{\mbox{\boldmath $\varrho$}}
\newcommand{\beps}{\mbox{\boldmath $\varepsilon$}}
\newcommand{\bxi}{\mbox{\boldmath $\xi$}}
\newcommand{\bbeta}{\mbox{\boldmath $\beta$}}
\newcommand{\be}{\begin{equation}}
\newcommand{\ee}{\end{equation}}
\newcommand{\bea}{\begin{eqnarray}}
\newcommand{\eea}{\end{eqnarray}}

\newcommand{\ii}{\mbox{i}}
\newcommand{\e}{\mbox{e}}
\newcommand{\pa}{\partial}
\newcommand{\Om}{\Omega}
\newcommand{\vep}{\varepsilon}
\newcommand{\bfph}{{\bf \phi}}
\newcommand{\lm}{\lambda}
\def\theequation{\arabic{equation}}
\renewcommand{\thefootnote}{\fnsymbol{footnote}}
\newcommand{\re}[1]{(\ref{#1})}
\newcommand{\R}{{\rm I \hspace{-0.52ex} R}}
\newcommand{\N}{{\sf N\hspace*{-1.0ex}\rule{0.15ex}%
{1.3ex}\hspace*{1.0ex}}}
\newcommand{\Q}{{\sf Q\hspace*{-1.1ex}\rule{0.15ex}%
{1.5ex}\hspace*{1.1ex}}}
\newcommand{\C}{{\sf C\hspace*{-0.9ex}\rule{0.15ex}%
{1.3ex}\hspace*{0.9ex}}}
\newcommand{\eins}{1\hspace{-0.56ex}{\rm I}}
\renewcommand{\thefootnote}{\arabic{footnote}}
 \maketitle
\begin{abstract} 
Different types of gravitating compact objects occuring in d=5 space-time are considered: boson stars, hairy black holes
and perfect fluid solutions.  
All these solutions of the Einstein equations coupled to matter
have well established counterparts in d=4; in particular neutron stars can be modell§ed 
more or less realistically by a perfect fluid.
A special emphasis is set on the possibility -and/or the necessity- for these solutions to have an intrinsic angular
momentum or spin. The influence of a cosmological constant on their pattern is also studied.
 Several physical properties  are presented from which 
 common features to boson  and neutron stars clearly emerge. 
 We finally  point out qualitative differences of the gravitational interaction supporting these
 classical lumps between four and five dimensions.   
\end{abstract}

\section{Introduction}

The theory of gravity proposed by Einstein a century ago is among the most successful models in theoretical physics. In particular, the 100 years old prediction of gravitational wave emission by compact binary systems has finally been observed in a direct way - two times - in 2015 by the gravitational wave observatories of the LIGO project. These historical events, GW150914 and the Christmas present GW151226 \cite{Abbott:2016blz, Abbott:2016nmj} constitute the first real strong field test of Einstein's gravity. Yet this successful theory has frustrating features and is not able to fully explain all of (cosmological) observations, or even galactic dynamics without introducing some unknown type of dark matter and/or an unexplained cosmological constant. On the other hand, from a theoretical perspective, the theory of gravity sits aside of the other 3 fundamental interactions, and is not compatible with a quantum description. 

Compact objects such as black holes and neutron stars are studied in order to test gravity or what is beyond theoretically. Black holes are clean objects, in the sense that they are solutions of the vacuum equations, but they pudicly hide a singular point where general relativity fails. On the contrary, neutron stars display no regularity problem, but are constituted by matter that is not well understood \cite{Lattimer:2000nx} because of the very high density regime inside these objects. Many models describing the microphysics inside a neutron star are available on the market, some of them being incompatible with some current observations, and still many of them being plausible candidates. 

Such compact objects can be used as theoretical laboratories to test alternative models of gravity. There have been many studies in the last years constructing neutron star models in, say scalar-tensor theories, Galileon theories  Eddington-inspired Born-Infeld theories, stringy gravity, and so on (see for instance \cite{Kleihaus:2016dui, Cisterna:2015yla, Pani:2011mg}, or \cite{Berti:2015itd} for a review).

Perhaps the simplest extension of pure gravity to a gravity-matter system
consist to couple (minimally or not) a scalar field to the Einstein-Hilbert action. In this system, boson stars
solutions can be constructed. 
These are solitonic, compact regular equations characterized by a conserved quantity. 
For a review of boson stars, see e.g.  \cite{Mielke:1997re}.
Recently a new class of hairy black holes was constructed \cite{Herdeiro:2014goa}. Interestingly
these black holes need to spin sufficiently fast and their domain of existence
is limited by a family of boson stars; this constitutes a new motivations for studying boson stars.

At the other side of the gravity community, higher dimensional models of gravity has been intensively studied. The main motivation being provided by string theory and the famous gauge/gravity duality. In this context, spacetimes with a (negative) cosmological constant is of particular relevance; stability of the $AdS$ spacetime has been debated \cite{Bizon:2011gg}, and more recently, dynamical process has started being investigated more seriously \cite{Rocha:2014gza, Delsate:2014iia}. The prototype vacuum higher dimensional solution is the Myers-Perry black hole \cite{MP}. It played an important role in understanding the dependence of the gravitational interaction to the number of  dimensions of space-time.  
Besides these solutions which present an event horizon and an essential singularity at the origin, boson star solutions can be constructed as well by supplementing gravity with an appropriate matter field. The first construction of this type was reported in \cite{Astefanesei:2003qy, Prikas:2004fx}. Enforcing the d-dimensional boson star to spin requires an appropriate choice of the boson fields as shown in \cite{Hartmann:2010pm}. The counterpart of the black holes of \cite{Herdeiro:2014goa} to five dimensions was achieved in \cite{Brihaye:2014nba} in the case of equal angular momenta and in \cite{Herdeiro:2015kha} for arbitrary angular momenta. The patterns of hairy black holes in d=4 and in d=5 present similitudes but the main difference is that the underlying boson stars and hairy black holes display a mass gap in d=5.

In this paper, we focus on higher dimensional (5-dimensional) self-gravitating perfect fluid. Contrary to the case of 4 spacetime dimensions, a non-perturbative analysis of rotations is technically much simpler, in the case where the two independant angular momenta are equal. In this case, the spatial isometry is enhanced (in 5 dimensions) from $U(1)\times U(1)$ to $U(2) \equiv U(1)\times SU(2)$. As a consequence, it is possible to describe the nonperturbative rotation in the form of a co-homogeneity 1 problem, i.e. using ordinary differential equations (ODEs), whereas the less symmetric case or the case of 4 spacetime dimensions is described by partial differential equations (PDEs). In many papers (see namely \cite{Yazadjiev:2016pcb, Cisterna:2016vdx}) the effect of rotation on neutron stars is 
emphasized by considering slow rotations, allowing to still describe the problem in a simpler mathematical form (ODEs) than the nonperturbative case (PDEs). 

To our knowledge, the equations describing a gravitating perfect fluid in higher dimensions
has been studied only marginally \cite{deBoer:2009wk, Arsiwalla:2010bt}.
Recently in Ref. \cite{Bordbar:2015wva}, static neutron stars were emphasized in $d$ dimensional space-time
with the emphasis set on a positive cosmological constant.
One natural extension of this work is to include the effect of rotation of the fluid. In this paper we
attempted to address the effect of rotation both pertubatively (independant angular momenta) and non-perturbatively (equal angular momenta). 

The paper is organied as follow~: In Sect.2 we present the general framework, the ansatz for the metric
and recall the form of the Myers-Perry solutions. Sect. 3 is devoted to the spinning, compact objects
supported by bosonic matter (boson stars and black holes). The solutions composed of perfect fluid-type matter
are emphasized in Sect. 4. The effects of a cosmological constant on these latter is pointed out in Sect. 5.
Some extensions of these results are mentionned in Sect. 6.

\section{The model and equations}
We will consider the Einstein equations supplemented by matter~:
\be
\label{general_equations}
        G_{\mu}^{\nu} + \Lambda g_{\mu}^{\nu} = \frac{8 \pi G_d}{c^4} T_{\mu}^{\nu} \ \ \ , \ \ \ {\rm "Matter \ equations"}
\ee
for two types of matter field $T_{\mu}^{\nu}$ corresponding to boson star and to a perfect fluid respectively.
The form of the energy momentum tensors will be specified later as well as the "Matter equations".
In this formula, $\Lambda$ represents a cosmological constant and $G_d$ the $d$-dimensional Newton constant.

For both cases, we will parametrize the metric according to
 \begin{eqnarray}
\label{metric}
ds^2 & = & -b(r) dt^2 + \frac{1}{f(r)} dr^2 + g(r) d\theta^2 + h(r)\sin^2\theta \left(d\varphi_1 - 
w(r) dt\right)^2 + h(r) \cos^2\theta\left(d\varphi_2 -w(r)dt\right)^2 \nonumber \\
&+& 
\left(g(r)-h(r)\right) \sin^2\theta \cos^2\theta (d\varphi_1 - d\varphi_2)^2 \ ,
\end{eqnarray}
where $\theta$ runs from $0$ to $\pi/2$, while $\varphi_1$ and $\varphi_2$ are 
in the range $[0,2\pi]$.
The corresponding space-times  possess two rotation planes at $\theta=0$ and $\theta=\pi/2$ and the isometry
group is $\mathbb{R}\times U(2)$.
The metric above still leaves the diffeomorphisms related to the definitions of the radial variable $R$ unfixed; 
for the numerical construction, we will fix this freedom by choosing $g(r)=r^2$. 

In the following, the matter fields will be choosen and parametrized
in such a way that the general equations (\ref{general_equations}) reduces consistently to a system of differential
equations in the functions $f,b,h,w$ and the functions parametrized the matter.

\subsection{Vacuum solutions: Myers-Perry black holes}
 Before setting the matter fields, we find it useful -for completeness- to recall
 the form of the asymptotically flat, vacuum solutions i.e. 
 for $\Lambda=0$ and  $T_{\mu}^{\nu}=0$ in (\ref{general_equations}). 
 These are the Myers-Perry solutions \cite{MP};
 they are spinning black holes with event horizon $R_H$ and horizon angular velocity $\Omega_H$. They have the form
 \bea
 f(r) &=& 1 - \frac{1}{1 - r_H^2 \Omega_H^2} \Bigl( \frac{r_H}{R} \Bigr)^2 + \frac{r_H^2 \Omega_H^2}{1 - r_H^2 \Omega_H^2} \Bigl( \frac{r_H}{r} \Bigr)^4 \nonumber \\
 b(r) &=& 1 - \Bigl( \frac{r_H}{r} \Bigr)^2 \frac{1}{1 - (1 - (\frac{r_H}{r})^4)r_H^2 \Omega_H^2 } \nonumber \\
 h(r) &=&  r^2 \Bigl( 1 +  \frac{r_H^2 \Omega_H^2}{1 - r_H^2 \Omega_H^2} \Bigl( \frac{r_H}{r} \Bigr)^4 \Bigr) \nonumber \\
 w(r) &=&  \frac{\Omega_H}{1 - (1 - (\frac{r_H}{r})^4)r_H^2 \Omega_H^2   }       \Bigl( \frac{r_H}{r} \Bigr)^4
 \eea
 It is well known that generic Myers-Perry solutions with an outside (event horizon) $r_H$
 presents a second horizon at $r'_H$ with $0 < r'_H < r_H$.
 The solutions exist for $r \in ]0, \infty[$; the functions $b(r), w(r)$ 
 remain finite in the limit $r \to 0$ while $f(r), h(r)$ diverge in this limit. 
 The solution present a singularity at the origin.
 
 \section{Boson stars}
 Boson stars are obtained by coupling scalar fields to the Einstein-Hilbert gravity. That is to say by using
 the action
 \be
\label{egbbs}
   S = \frac{1}{16 \pi G_5} \int d^5 x \sqrt{- g} ( R - 2 \Lambda 
   - (16 \pi G_5) ( \partial_{\mu} \Pi^{\dagger} \partial^{\mu} \Pi + M_0^2 \Pi^{\dagger} \Pi ) )
\ee
Here $R$ represents the Ricci scalar, $\Lambda=-6/\ell^2$ is the cosmological constant.
The matter sector  consists of a doublet of complex scalar fields with the same mass $M_0$
and denoted by $\Pi$ in (\ref{egbbs}). The corresponding form of the energy momentum tensor is obtained
in the standard way~:
\be
T_{\mu \nu} = \Pi_{\mu}^{\dagger} \Pi_{\mu} + \Pi_{\nu}^{\dagger} \Pi_{\mu} - g_{\mu \nu}
\Bigl[ \frac{1}{2} g^{\alpha \beta}(\Pi_{\alpha}^{\dagger} \Pi_{\beta} + \Pi_{\beta}^{\dagger} \Pi_{\alpha} ) + M_0^2 \Pi^{\dagger} \Pi   \Bigr]
\ee

The key point for the construction of classical solutions is to choose the scalar doublet in the form 
\be
\label{harmonic}
          \Pi(x) = \phi(r) e^{i \omega t} \hat \Pi
\ee
where $\hat \Pi$ is a doublet of unit length that depends on the angular coordinates.
The standard non spinning solutions are recovered by means of the particular form 
\begin{equation}
\label{hatphi_non}
 \hat \Pi = (1,0)^t  \ .
\end{equation}
Only one component of the scalar doublet  is non-zero. 
This leads  to a system of coupled equations for the fields $f(r), b(r), \phi(r)$
while $h(r)=r^2$, $w(r)=0$.
In contrast, spinning solutions can be obtained with the  parametrization  \cite{Hartmann:2010pm}
\be
\label{hatphi_rot}
\hat \Pi = \left(\sin \theta e^{i \varphi_1},\cos \theta e^{i \varphi_2}  \right)^t \ .
\ee  
 All fields $f(r),b(r),h(r),w(r)$ and $\phi(r)$ corresponding  Einstein-Klein-Gordon  equations 
 are then non-trivial and the full system  admit regular, localized solutions~: 
 the spinning boson stars. 
 They  can be characterized by several physical parameters, some of them are associated with the globally 
conserved quantities. The charge $Q$ associated to the U(1) symmetry possesses a special relevance;
in terms of the ansatz (\ref{metric})
it takes respectively the forms
\begin{eqnarray} 
  Q = - \int \sqrt{- g} j^0 d^4 x &=& 4 \pi^2 \int_0^{\infty} \sqrt{\frac{bh}{f}} \frac{r^2}{b} (\omega + w) \phi^2 d r 
  \ \ , \ \ j^{\mu} = -i (\Pi^{\dagger} \partial^{\mu} \Pi -  \partial^{\mu} \Pi^{\dagger} \Phi) \ .
\end{eqnarray}
 The mass of the solution $M$
and the angular momentum $J$ can be obtained from the asymptotic decay of some
components of the metric~:
\begin{equation} 
                      g_{tt} = -1 + \frac{8 G_5 M}{3 \pi  r^2} + o(\frac{1}{r^3}) \ , \
                      g_{\varphi_1 t} = - \frac{4 G_5 J}{\pi r^2} \sin^2 \theta + o(\frac{1}{r^3}) \ , \
                      g_{\varphi_2 t} = - \frac{4 G_5 J}{\pi r^2} \cos^2 \theta + o(\frac{1}{r^3})
\end{equation}
In the case of spinning solutions, 
it was shown \cite{Hartmann:2010pm} that the charge $Q$ is related to the sum of the two (equal) angular
momentum of the solution~: $Q = 2 |J|$.

Note that the $5$-dimensional Newton constant has units $[G_5] = length^2 mass^{-1}$.
   We will adopt the dimensions such that $8 \pi G_5 = 1$ and  $c=1$. As a consequence, all physical
   dimensions can be expressed in the unit of a length; in particular $[M_0] = length^{-1}$. 
   The following dimensionless quantities will be used on the plots to characterize the solutions
   \be
       x = r M_0 \ \ , \ \ \omega M_0^{-1}    \ \ , \ \ Q M_0 \ \ , \ \ M M_0^2
   \ee 

For both cases -spinning and non-spinning- the system of differential equations 
has to be supplemented by boundary conditions. The regularity of the equations at  
 the origin requires
\be
      f(0) = 1 \ \ , \ \ b'(0)= 0 \ \ , \ \ h'(0) = 0 \ \ , \ \ w(0) = 0 
\ee
and $\phi'(0)=0$ (resp. $\phi(0) = 0$) in the spinning (resp. non-spinning) case.
Then, assuming for the moment $\Lambda =0$ and  the metric to be asymptotically flat, localized solutions  
should obey the following conditions~:
 \be
    f(r \to \infty) = 1 \ \ , \    b(r \to \infty) = 1 \ \ , \ \ \frac{h}{r^2}(r \to \infty) = 1 \ \ , 
       \ \ w(r \to \infty) = 0 \ \  , \ \ \phi(r\to \infty) = 0 .
 \ee

The construction of exact solutions obeying the equations-plus-boundary conditions can be achieved 
numerically; for this purpose, we used the routine COLSYS \cite{colsys}.
In principle, the frequency $\omega$ is  fixed by hand and the value $\phi(0)$ (or $\phi'(0)$)
is fine tuned up to an $\omega$-depending value such that all boundary conditions are obeyed.
Alternatively, it is also convenient to use $\phi(0)$ (or $\phi'(0)$) as control parameter and to
reconstruct the corresponding frequency by an appropriate reformulation of the equations. 

\begin{figure}[ht!]
\begin{center}
{\label{fig_0}\includegraphics[width=12cm]{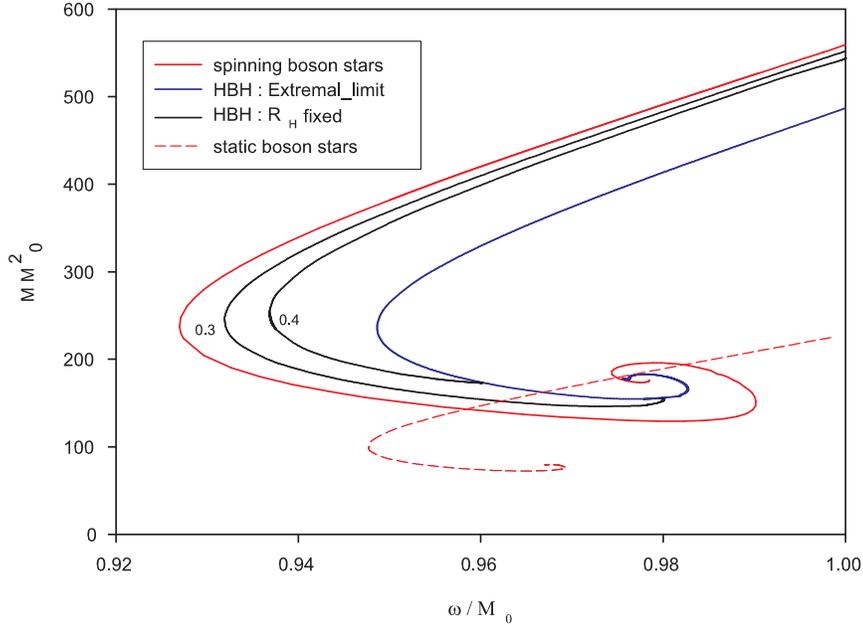}}
\end{center}
\caption{Mass of the non spinning  and spinning bosons stars as function of $\omega$ (solid and dashed red lines).
The mass of spinning black holes corresponding to horizon $x_h=0.3$ and $x_h=0.4$ is  shown by the black lines. 
The blue line represent the limit where black holes become extremal. 
\label{mass_omega}
}
\end{figure}

The results show that spinning and non-spinning solutions exist for a finite interval of $\omega$; the 
plot of the mass as function of $\omega$ is shown on Fig. \ref{mass_omega}.
The solid and dashed RED curves represent respectively the mass of the spinning and non-spinning boson stars.
The limit $\omega \to 1$ correspond to $\phi(0) \to 0$ (non-spinning case) and $\phi'(0) = 0$ (spinning case).
In fact in this limit the scalar fields tends uniformly to zero; interestingly the mass of the solution remain
finite in this limit, demonstrating the occurence of a mass gap \cite{Brihaye:2014nba}.
This seems to be a peculiarity of the gravitational
interaction in $d=5$. 

While increasing the control parameter $\phi(0)$ (or $\phi'(0)$) the frequency $\omega$ decreasing up to
a minimal value, say $\omega_{m}$ (for instance we find respectively $\omega_m \approx 0.925$ and 
$\omega_m \approx 0.947$ for the spinning and non-spinning cases). increasing again the control parameter
a succession of branches are produced which exist on smaller, nested intervals of $\omega$.
The plot of the mass versus frequency then presents the form of a spiral. For simplicity
we do not report the value of the charge $Q$ but the shape of the curves are qualitatively similar.

\begin{figure}[ht!]
\begin{center}
 {\label{fig_0}\includegraphics[width=12cm]{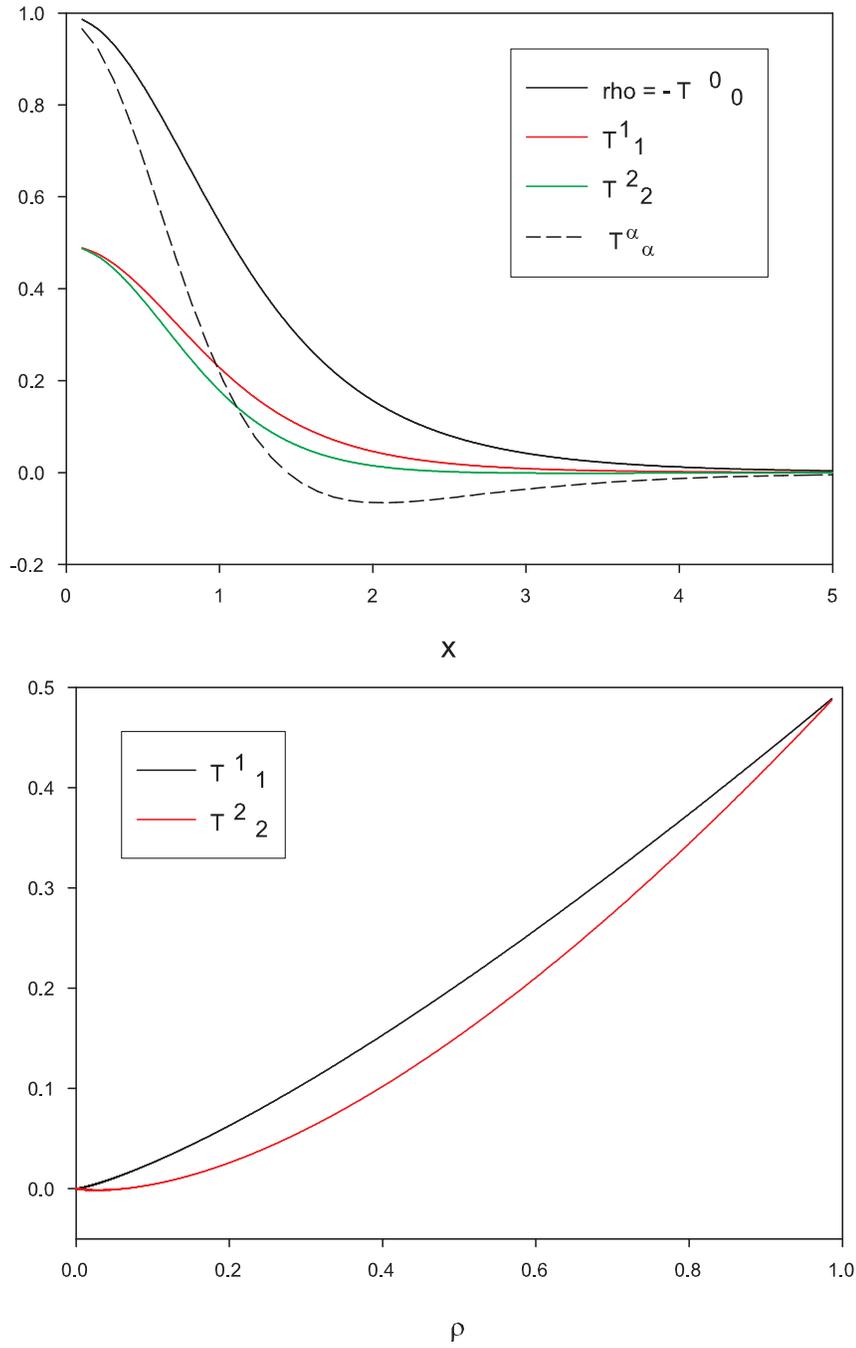}}
\end{center}
\caption{Up: The components of the energy momentum tensor of the non spinning boson star with $\phi(0)=0.5$
as function of $x$. Bottom: The pressure as function of the density . 
\label{energy_momentum}
}
\end{figure}

In the next section
we will consider matter under the form of a perfect fluid for which specific
relations are assumed for the different components of the energy-moentum tensor.
In a purpose of comparison , we  show
the different components of the energy momentum tensor of a boson star
on Fig. \ref{energy_momentum}. This solution is generic, corresponding to $\phi(0)=0.5$ 
and $\omega = 0.948$. The pressure-density relation is also provided in the lower figure.
 
\subsection{Black holes} 
In \cite{Brihaye:2014nba}, it was demonstrated that  the spinning boson stars can be deformed into 
spinning hairy black holes.
Within the ansatz (\ref{metric}) used for the metric the black holes are characterized by an event horizon
at $x = x_H$ and by the conditions $f(x_H)=0, b(x_H) = 0$.  
We refrain to discuss the different conditions
of regularity that the different functions  should fullfill at the horizon, they are discussed at length in \cite{Brihaye:2014nba} and \cite{Brihaye:2016vkv}. 
For brevity, we just mention the one of them -essential for the discussion- namely~: 
\be
w(x_H) = \omega \ \ .
\ee
Accordingly, the hairy black holes can be labelled by 
two parameters~: the horizon $x_H$ and the angular velocity on the horizon $w(x_H)$. 
A sketch of the
 pattern of hairy blach holes is given  by Fig. \ref{mass_omega}, the masses of these black holes are
 set between the red lines (corresponding to bosons stars) and the blue lines (corresponding to extremal black holes).
 The mass-frequency relation for the black holes with $x_H=0.3$ and $x_H=0.4$ are represented by the black lines.

\clearpage
 \section{Perfect Fluid}
 In this section, we  study perfect fluid-like compact objects  in five dimensional gravity
 and  compare their   features with the boson star-solutions constructed in the previous section.
 The idea is to solve the Einstein Equations by taking the energy momentum of a perfect fluid;, i.e.
 \be
           T^{\mu \nu} = (\rho + P) U^{\mu} U^{\nu} + P g^{\mu \nu}   \ \ , \ \ U^{\mu}U_{\mu}= -1. 
 \ee
Here $\rho, P$ represent the density and the pressure which we both assume to depend on $r$ only and
$U^{\mu}$ is the four velocity of the fluid.  Assuming $U^1 = U^2 = 0$ and $U^4=U^3$, 
the normalization condition
implies a relation between $U^0$ and  $U_4$.
Referring to the case $d=4$, we further pose $U^4(r) = \lambda \omega(r)/ \sqrt{b(r)}$ where $\lambda$
is a constant. Then the normalization condition leads to the relation
 \be
         U^0 = \frac{\lambda h \omega^2 \pm \sqrt{b^2 + (\lambda^2-1) b h \omega^2}  }{\sqrt{b}(b - h \omega^2)}
 \ee
So it exists a continuum of possible values for $U$ (this contrasts to the case d=4 where the normalisation
leads to a unique solution).
 In the following, we will only study the two cases corresponding to $\lambda^2 = 1$.
 The relation between $U^0$ and $U^4$ then simplifies~:
  \be
      U^0 = \epsilon \frac{1}{\sqrt b}\ \ , \ \ U^3 = U^4 = \frac{\epsilon w}{\sqrt{b}} \ \ \ , 
      \ \ \epsilon = \pm 1 \ \ \  \ \ {\rm : Type \ I}
 \ee
 or
 \be
      U^0 = -  \frac{\epsilon}{\sqrt{b}} \ \frac{b + h w^2}{b - h w^2} \ \ , \ \ U^3 = U^4 = \frac{\epsilon w}{\sqrt{b}}
      \ \   \ \ {\rm: Type \ II}
 \ee
 
\subsection{Static fluid} 
Assume  a static fluid with $w=0$, the equations corresponding to $w$ and $h$ are trivially solved by
$h(r)=r^2$, $w(r)=0$ and 
the relevant Einstein equations are those of  
 $f$ and $b$ and are completed by
the conservation equation for the perfect fluid~:   
\be
               P' + \frac{b'}{2 b} (P + \rho) = 0 \ .
\ee 
We further impose an equation of state 
between $\rho$ and $P$ under the form of a polytrope ~: $P = K \rho^{1 + 1/n}$,
where $K$ and $n$  are constants.  For definiteness we set $n=2$.
For later use, we introduce the function $Z(r)$ such that $\rho = Z^n$, $P = K Z^{n+1}$.

Using an appropriate rescaling of the density function, it turns out that the system depends
on the coupling parameters $G_5$ and $K$ through the combination   $\kappa \equiv 8 \pi G_5 K^{-n}$. 
The effective parameter $\kappa$  can then be absorbed in a rescaling of the radial variable $r$. 
'Physical' quantities can be recovered by this dimensionless system by noticing that 
the radius $R$ and the mass $M$ of the star scale respectively according to $K^{-n/2}R$ and $K^n M$.   

\begin{figure}[ht!]
\begin{center}
{\label{fig_0}\includegraphics[width=8cm]{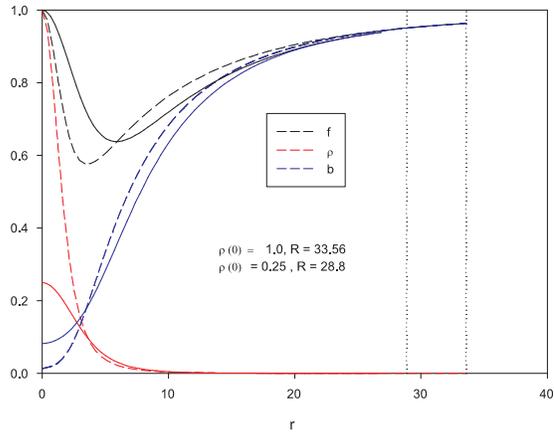}}
\end{center}
\caption{Metric function $f$ and density for type I 
perfect fluid.
\label{fig_0}
}
\end{figure}  

The Einstein-perfect-fluid equations lead to a system of three coupled equations for $f,b,P$. 
The equations for $f(r)$ and $P(r)$ are of the
first order while the equation for $b(r)$ is of the second order. Four boundary conditions
are then necessary to specify a solution. 
To obtain the solution with a given value of the radius, say $R$, 
the system is first solved on  the interval $r \in [0,R]$.
The requirements for the metric to be  regular at the origin and for the matter to vanish for $r \geq R$
(then ensuing a Schwarschild-Myers-Perry metric in this region) lead to
\be
     f(0) = 1  \ , \ b'(0) = 0 \ \ , \ \ f(R) = b(R) \ \ , \ \ P(R) = 0 \ \ .
\ee  
In the outside region $r \in [R, \infty]$, the solution is matched by continuity
with the appropriate Schwarschild-Myers-Perry solution.

 In principle the radius $R$ is a natural parameter to control the solutions but it turns out
 that varying $\rho(0)$, the density at the center, is slightly more convenient for the construction~:
  the radius $R$ can then be identified as the first zero of $P(r)$.
  Profiles of interior solutions corresponding to $\rho(0) = 1$ and $\rho(0) = 0.25$ are reported in Fig. \ref{fig_0};
the corresponding radius are given in the caption.
 
 \begin{figure}[ht!]
\begin{center}
{\label{fig_0}\includegraphics[width=12cm]{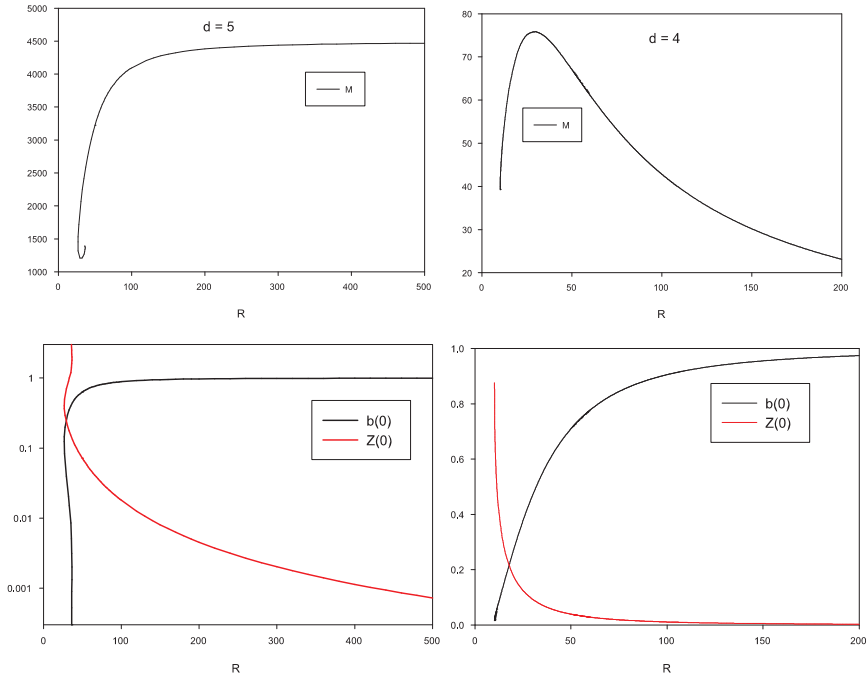}}
\end{center}
\caption{Comparaison of the mass and  quantities $b(0)$ and $Z(0)$ for non spinning perfect fluid for $d=4,5$. 
\label{comparaison}
}
\end{figure}    
Several parameters characterizing the d=5 perfect fluids are presented on Fig. \ref{comparaison} (left panels)
and compared with the corresponding data available for the more known d=4 case (right panels).

The mass $M$ as function of the radius $R$ is reported on the upper part of the figure. 
The parameters $b(0)$ and $Z(0)$ (with $Z(r) \equiv P^2(r)$) are also given on the lower panel of the figure. 
We see that solutions exist for large enough values of the radius, say for $R > R_{min}$.
Our numerical results  reveal the existence of two branches of solutions. The main branch exists for 
$R \in [R_{min}, \infty]$ while the second branch exists  
 for $R \in [R_{min}, R_c] \sim [26.8, 36.6]$. For the values of $R$ such that two solution exist,
 the solution with the lowest mass belong to the second branch. In the limit $R \to R_{min}$, the two masses coincide. 
The solutions the main branch exist for large values of $R$. The limit $R \to \infty$ in fact corresponds to
the density and pressure tending to zero and the solution smoothly approaching Minkowski space-time. 
Along with the case of boson stars, the mass of the star does not approach zero in the limit of vanishing matter~:
both, boson stars and perfect fluid solutions present a mass gap.

We now comment on the critical phenomenon limiting the solutions of the second branch of solutions
for $R \to R_c$. It turns out that this limit is approached for $\rho(0) \gg 1$,  at the same time 
the metric parameter $b(0)$ tends to zero. 
This suggests that, taking the limit $\rho(0) \to \infty$, 
the metric becomes singular at the origin and that the mass of the limiting configuration remains finite.
In fact, this phenomenon is qualitatively the same for the boson stars, see Fig.\ref{fig_0}.

Owing these peculiar properties of the d=5 perfect fluid, 
it is natural to compare these solutions with the
more conventional  d=4 case. The corresponding data are reported respectively on the left and right
panels of Fig. \ref{comparaison}.
On the mass-radius plot, the main difference is that, for large $R$  the mass presents a local maximum
for $d=4$ while it keeps motononically increasing with $R$ for $d=5$.
In both cases, large values of $R$ corresponds to a vanishing of the perfect fluid energy momentum tensor
and to a metric approaching d-dimensional Minkowski space-time.
However only for $d=5$ the solutions present a mass gap.

For the small values of $R$ the d=4-pattern is simpler, presenting only one branch. 
The solution exist for $R \geq R_{min}$ with $R_{min} \approx 9.5$. In the limit $R \to R_{min}$
the central density diverges and the metric parameter $b(0)$ approaches zero.

\subsection{Type I spinning fluid}
We now investigate the possibility of a spinning perfect fluid by solving the full equations 
with a spinning perfect fluid 
 as constructed at the top of this section.
For type I the conservation law reads as above 
\be
               P' + \frac{b'}{2 b} (P + \rho) = 0 \ \ ,
\ee 
and is supplemented by the polytropic equation of state.
The problem consists in  integrating the full Einstein equations  
as an initial value problem from the origin up to some finite value of $r$ with the conditions
\be
  f(0) = 1 \ , b(0) = 1 \ , \ b'(0) = 0 \ , \ h(0) = 0 \ \ , \ h'(0) = 0 \ \ , \ w(0) = \Omega \  \ , \ w'(0) = 0 \ .
\ee
We keep in mind that the metric field $b$ can be renormalized at the end 
in order to obey $b(\infty)=1$.
 These conditions  are  completed by 
 \be
          \rho(0) = \rho_0 \ \ , \ \ \rho'(0) = 0 \ \ .
 \ee
 Integrating the system up to a large enough value of $R$, the goal was to 
 look for the condition $\rho(R_S) = 0$ we could determine the radius of the star $R_S$
 as a function of $\Omega$, $\rho_0$ and the constant $K$.
 
 Unfortunately, the Einstein equations for the fields $\omega$ and $h$
 are trivially fulfilled with $\omega(r) = \Omega$
 and $h(r) = r^2$. 
 In particular, the equation for $w(r)$  has the form
 \be
           w'' = w'( \frac{5H-12f-8}{3 x f} + \kappa \frac{x \rho}{f} + "lower \ order" ) \ \ , \ \ H(x) \equiv \frac{h}{x^2}
 \ee 
 We see that the most singular term implies $w(r) = \omega_0 + \frac{C}{r^4}$, excluding the possibility 
 of non-trivial, regular solutions at the origin.
 This contrasts with the corresponding  equations for d=4. 
 In the equation above, we have checked that the term due to matter does not regularize the singular
 term. It is instructive to compare the above equation with its "boson-star" counterpart~: 
 \be
       w'' = w'( \frac{5H-12f-8}{3 x f} + "lower \ order" ) + 2 \alpha \frac{\phi^2}{x^2 f H}(\omega + w) \ , \ \ 
       \alpha \equiv 8 \pi G_5 M_0^2 \ .
 \ee
 Here  the contribution of matter can be used to suppress the singular term, leaving the possibility of 
 existence of spinning
 solutions which were indeed found  numerically.

\subsection{Independant slow rotations}
In light of the results discussed in the previous subsections, we have checked whether it was possible to excite a single rotation, or non equal independant rotations. The underlying hypothesis to test being that the fluid cannot be made rotating due to the symmetry enhancement of the equal angular momentum spacetime.

 The anzatz for the metric is proposed  in \cite{Herdeiro:2015kha} and leads to a system
 of partial differential equations with boundary conditions. 
 Due to the complexity of these equations, we limited the analysis
 to small rotation and linearized the equations in the two independant rotating functions. 
For this purpose, we have followed the standard procedure for computing the slow rotation equations to order 1 \cite{hartle} namely, we considered a spherically symmetric background and added the rotations as first order perturbations around the spherically symmetric case.

For completeness, we present here the form of the metric truncated to first order in rotation:
\be
ds^2 = -b(r) dt^2 + \frac{dr^2}{f(r)} + g(r)\left(  d\theta^2 + \sin^2\theta (d\varphi - \epsilon \omega_1(r) dt)^2 + \cos^2\theta (d\psi - \epsilon \omega_2(r)dt)^2  \right),
\ee
where $\omega_1, \omega_2$ are the independant angular velocity functions, $b,f$ are the spherically symmetric background functions, $g$ is an arbitrary function (as above we set  $g=r^2$) and $\epsilon$ is the small parameter controlling the slow rotation approximation.

The matter fields are still modelled by a perfect fluid of the form $T_\mu{}^\nu = (\rho + P )u_\mu u^\nu + P \delta_\mu{}^\nu$, where $u$ is such that $u^2 = -1$, at least to first order in $\epsilon$, leading to 
$u^a =\frac{1}{\sqrt{b}}(1,0,\epsilon \omega_1, \epsilon \omega_2)$.

We find two additional equations for the functions $\omega_1,\ \omega_2$, given by
\be
\omega_i ''  -\left(\frac{\kappa  r P(r)}{3 f(r)}+\frac{\kappa  r \rho (r)}{3 f(r)}-\frac{5}{r}\right) \omega_i' = 0,
\label{eq:slowrot}
\ee
where $i=1, 2$. Note that the equations are the same for $\omega_1$ and $\omega_2$; only the initial conditions makes the difference.

The vacuum solution is given by
\be
\omega_i = \Omega_*\left( 1 - \frac{I}{r^4}\right), 
\ee
where $\Omega_*, I$ are constants of integration related to the rotation frequency of the configuration, measured by a distant observer, and to the moment of inertia of the configuration.

For a solution to be regular at the origin, the boundary conditions for $\omega_i$ are 
\be
\omega_i(0) = \Omega_i,\ \omega_i'(0) = 0,
\ee
for constant values $\Omega_i$.

Due to the fact that there are no terms linear in $\omega_i$, and no source term in equation \eqref{eq:slowrot}, the only regular solutions are $\omega_i = \Omega_i \in \mathbb R$, as in the equal angular momenta case studied in the previous section.

Here again, five dimensinal spacetime cannot support perfect fluids in rotation, even in the case of unequal angular momenta.

\subsection{Type II fluid} 
For type II the conservation equation is more involved and not worth to be written explicitely.
From the beginning it is clear  the condition $b - h w^2 > 0$ will limitate the radius of the star. 
We were able to we construct numerically solutions with $dw/dr \neq 0$ by using a shooting method from the origin. 
However these solutions obtained for $w(0) > 0$  are hard to be interpreted for several reasons, namely~:
(i) after decreasing  the density function reaches a local minimum and then increases;
(ii) the functions $w(r),b(r), \rho(r)$ have a tendency to diverge for a finite radius while $f(r)$
approaches zero.
These results are illustrated by mean of Fig. \ref{fig_1}; the features of this graphic seem to be generic.

In summary, from the numerous cases studied, the result strongly suggest that the d=5 perfect fluid
cannot be made spinning.

   \begin{figure}[ht!]
\begin{center}
{\label{fig_1}\includegraphics[width=8cm]{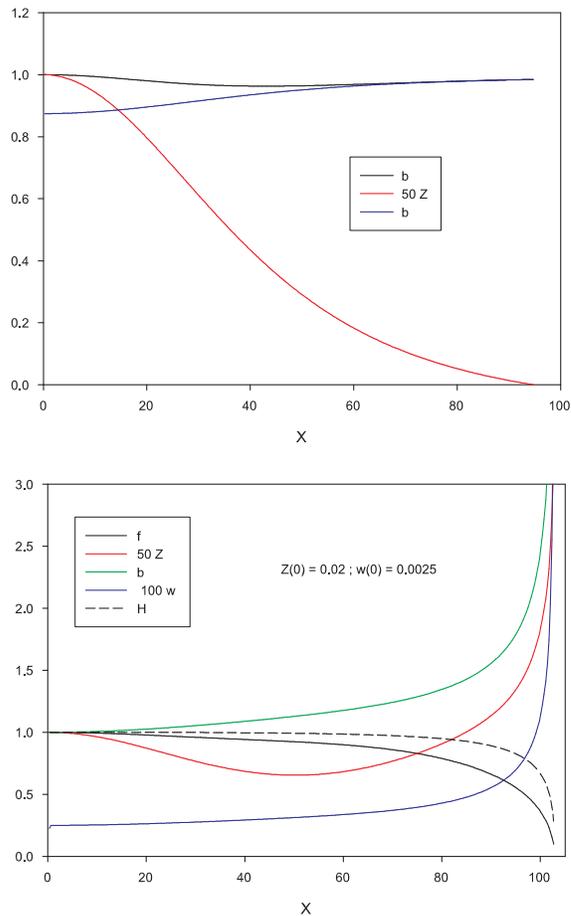}}
\end{center}
\caption{Up: Metric functions $f,b, Z$ for the non spinning perfect fluid with $Z(0)=0.02$.
Bottom: The corresponding solution with $w(0) = 0.0025$ 
and density for type II 
perfect fluid. A solution with $w \neq 0$ is superposed.
\label{fig_1}
}
\end{figure}  

 \section{Effect of a cosmological constant}
 
 In the above section, we obtained several characteristics of the aymptotically flat
 boson stars and perfect fluid solutions in 5-dimensions. In this section we address the 
 effect of a cosmological constant on these families of solutions. 
 Asymptotically AdS spinning boson stars (and black holes) have been studied in details in 
 \cite{Dias:2011at}. A crucial difference is that the presence of the negative cosmological
 constant suppress the mass gap in the pattern of solutions. On the other hand, the results
 of \cite{Brihaye:2016vkv} seem to exclude the existence of asymptotically deSitter spinning boson stars
 and black holes. 
 
\begin{figure}[ht!]
\begin{center}
{\label{fig_1}\includegraphics[width=8cm]{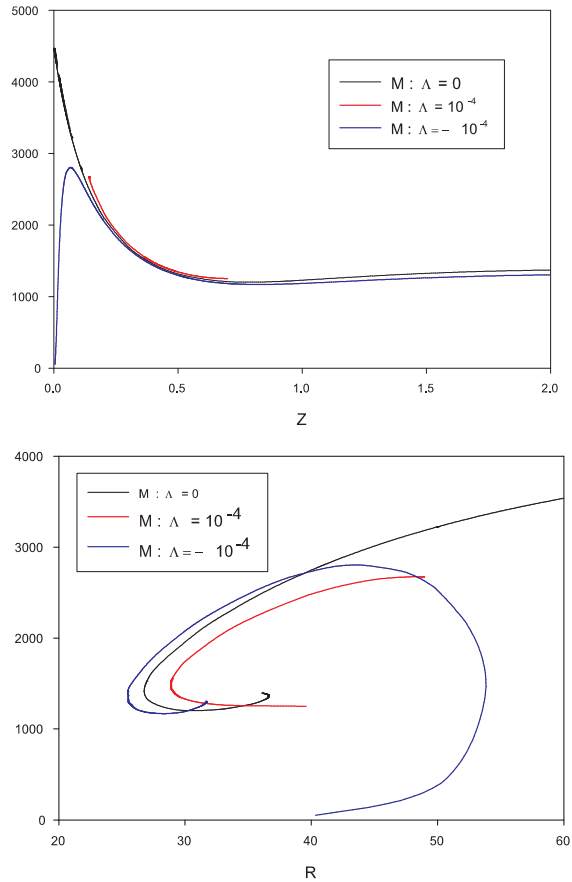}}
\end{center}
\caption{Several properties of d=5 solutions for $\Lambda=0, \pm 10^{-4}$.
\label{compa_lambda}
}
\end{figure}

 We therefore put the emphasis on the deformation of the static
 perfect fluid solutions by a cosmological constant. 
  We checked that the following properties hold for several values of $\lambda$,
 although -for clarity- the data on Fig.\ref{compa_lambda} is reported for $\Lambda = \pm 10^{-4}$.

 {\bf $\Lambda > 0$} We solved the equations in the interior in the case of a positive cosmological
 constant by integrating from the origin and using $\rho(0)$ as a shooting parameter.
 We observe that the function $Z(r)$ related to the density $\rho(r)$  crosses zero only for a finite range of
 the value $\rho(0)$ of the central
 density, say $\rho(0) \in [\rho_1, \rho_2]$.
 The maximal mass of the star is reached for $\rho(0) = \rho_2$; it coincides with the maximal value, say $R_{max}$ of the radius.
 This appears by means of the red curve of Fig.\ref{compa_lambda}. 
 The solutions exist for $R \in [R_{min}, R_{max}]$.
 Along the case $\Lambda=0$ two branches of solutions exist for small values of the radius.
 
 {\bf $\Lambda < 0$} In the case $\Lambda < 0$, the situation is quite different. 
 The following features are worth pointing out~:
 \begin{itemize}
 \item The solutions exist in the limit
 $\rho(0) \to 0$ and the mass naturally approaches zero in this limit (see blue line on Fig. \ref{compa_lambda}). 
 In other words, the presence negative
 cosmological constant suppresses the mass gap. It is remarquable that this feature also holds in the 
 case of AdS boson stars (see \cite{Dias:2011at}). The mass gap seems to be a characteristic
 of asymptotically flat d=5 compact objects.
  \item  As a consequence of the first property,
 the family of gravitating perfect-fluid presents 
 a configuration  with a maximal mass.  This is reached for a specific value of the radius say $R(\Lambda)$. 
 The value $R(\Lambda)$  tends to infinity while $\Lambda \to 0$.
 \item The solutions exist only for a finite interval of values of the radius, say $R \in [R_{min}, R_{max}]$.
 \item In the neighbourhood of both $R_{min}$ and $R_{max}$ there are two branches of solutions with two different
 masses corresponding to one value of $R$ .
 \end{itemize}
 
 Some of these features, but not all, agree with the results obtained in \cite{Bordbar:2015wva}. 
 The differences might be set on account
 on a different choice of the equation of state and/or on the accuracy of the numerical method.
 
\section{Outlook}
 Myers-Perry black holes provide a laboratory to test how the dimensionality of space-time affects the gravitational
 interaction on black holes and then to appreciate how special is the d=4 Kerr Black holes. Besides black holes several
 type of solutions have been constructed, namely by supplementing different sort of matter. 
 One of the most recent beeing the hairy
 black holes of \cite{Herdeiro:2014goa} which has to be supported by rotation.
 The recent interest for  gravity and field theory in higher-dimensional space-times motivates naturally the construction
 of boson stars, black holes as well as perfect fluid solutions in higher dimensions.
 This paper constitutes an attempt in this direction, although focalizing to d=4 and d=5 only. The generalisations and
 extentions are multiple but we mention only two~: (i) are spinning perfect fluid possible for $d > 5$; (ii) can perfect
 fluid surrounded by a cloud of boson field exist as stationnary solutions ?. 
  
{\bf Note added:} While completing the manuscript, Ref. \cite{Bhar:2016fne} appeared were five-dimensional compact
objects are emphasized as well.
 

\end{document}